\documentclass[aps,pra,twocolumn,groupedaddress,floatfix,showpacs]{revtex4}

\usepackage{graphicx}
\usepackage{textcomp}

\begin{document}

\title{~\vspace{1.5cm}\\ Optical torque controlled by elliptical polarization}


\author{M. E. J. Friese}

\author{T. A. Nieminen}
\email[]{timo@physics.uq.edu.au}

\author{N. R. Heckenberg}
\author{H. Rubinsztein-Dunlop}

\affiliation{Centre for Laser Science, Department of Physics,
The University of Queensland, Brisbane QLD 4072, Australia}


\date{4th August 1997}

\begin{abstract}
\vspace{-5cm}
\noindent
\hspace{-1.5cm}\textbf{Preprint of:}

\noindent
\hspace{-1.5cm}M. E. J. Friese, T. A. Nieminen, N. R. Heckenberg,
and H. Rubinsztein-Dunlop,

\noindent
\hspace{-1.5cm}``Optical torque controlled by elliptical polarization'',

\noindent
\hspace{-1.5cm}\textit{Optics Letters} \textbf{23}, 1--3 (1998)

\hrulefill

\vspace{3cm}

We show theoretically and demonstrate experimentally that highly absorbing
particles can be trapped and manipulated in a single highly focused
Gaussian beam. Our studies of the effects of polarized light on such
particles show that they can be set into rotation by elliptically
polarized light and that both the sense and the speed of their rotation
can be smoothly controlled.
\end{abstract}


\maketitle

Over the past 20 years the use of light to manipulate microscopic particles
has progressed from the more complicated multiple-beam radiation pressure
traps of Ashkin~\cite{ref1} and Roosen and Imbert~\cite{ref2} to simpler
single-beam traps. The best known of these is the single-beam gradient
optical trap (so-called optical tweezers), which can be used to manipulate
transparent highindex microscopic particles, or low-index particles if a
doughnut-shaped beam is used~\cite{ref3}. This trap is three dimensional
in the sense that, as well as experiencing a radial force, a particle
experiences an axial force that draws it toward the beam waist, allowing
it to be levitated even by a downward-propagating beam. The same experimental
arrangement can be used to trap metallic particles three dimensionally if
they are small enough to behave as dipoles~\cite{ref4}, but larger reflective
and absorbing particles experience too large a radiation pressure to be
levitated. However, these particles can be trapped radially against a
surface (i.e., a two-dimensional trap). Reflective metal particles have
been trapped in this way with a Gaussian beam~\cite{ref5}, but trapping
of micrometer-sized absorbing particles has been limited to traps that
use laser beams with a central field minimum, with a high-intensity ring
of light to confine the particles to a dark region in the center.

We show both theoretically and experimentally that strongly absorbing
particles can in fact be trapped and manipulated by radiation pressure
by use of a single Gaussian beam. Moreover, using a Gaussian mode provides
a unique opportunity to study the effects of the optical torque on absorbing
particles that are due to polarization alone, in contrast to our previous
work with an LG$_{03}$ doughnut mode~\cite{ref6,ref7} and that of
Simpson et al.~\cite{ref8} where torque owing to orbital angular momentum
was also present. Our experiments show that absorbing particles trapped
in a Gaussian beam are set into rotation by elliptically polarized light
and rotate in a direction that depends on the handedness of the ellipticity.
We also show that particles experience a torque that is due to elliptically
polarized light which is proportional to the angular momentum density of
the beam.

\begin{figure}[b]
\includegraphics[width=\columnwidth]{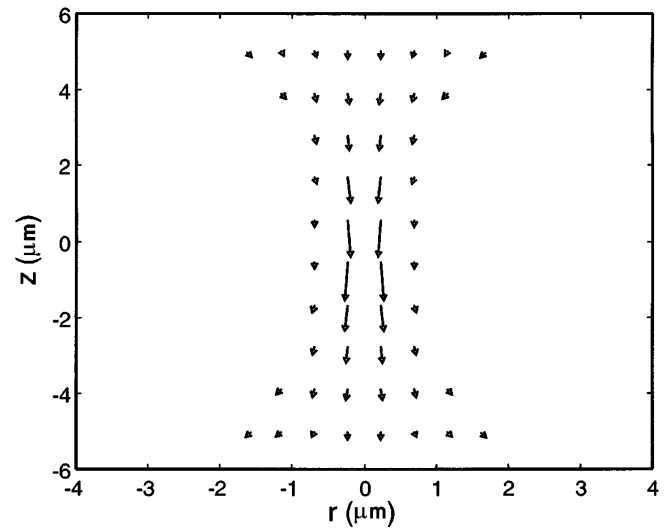}
\caption{Linear momentum in a focussed Gaussian laser beam. The direction
and magnitude of the Poynting vector above and below the beam waist are
shown. As the beam is converging, the Poynting vector has an inward radial
component at all points away from the beam axis, which permits
two-dimensional radial trapping of absorbing particles.}
\end{figure}

Two-dimensional trapping of absorbing particles can easily be understood
in terms of linear momentum transfer. The force $\mathbf{F}$ experienced
by a small area $\mathrm{d}A$ of an absorbing particle can be determined
from the time-averaged Poynting vector $\mathbf{S}$ by
$\mathbf{F} = (1/c) \mathbf{S}\cdot(-\mathrm{d}A)(\mathbf{S}/|\mathbf{S}|)$,
and the force on a particle can be obtained by integration over its surface.
Figure 1 shows the direction and magnitude of the Poynting vector above and
below the waist of a focused Gaussian beam. From this, we see that a particle
positioned above the waist will experience a force that has a component both
in the direction of beam propagation and radially inward, so the particle is
confined to the center of the beam. Conversely, a particle below the waist
experiences a force in the propagation direction and radially outward,
expelling it from the beam center. The radiation pressure force in the
direction of beam propagation can be countered by the normal reaction force
of the surface on which the particle is trapped. For a given radial trapping
force, the unwanted axial pressure will be greater for the Gaussian beam
than for a doughnut, increasing friction and causing trapping to be less
stable. However, particles trapped in this way can be manipulated, even
though these factors make the Gaussian beam trap a little harder to use.

The optical torque from elliptically polarized light, first calculated
in 1900 by Sadowsky~\cite{ref9}, was considered too small for experimental
detection until Beth's
famous experiment in 1936~\cite{ref10}, in which this tiny torque was
measured in a complicated and difficult experiment. With the additional
tools of the laser and the optical trap, it is now possible to observe
this torque acting on a microscopic scale~\cite{ref7,ref11} with relative
ease, as the effects are larger by several orders of magnitude. Optical
torque calculations by Marston and Crichton~\cite{ref12}, Chang and
Lee~\cite{ref13}, and Barton et al.~\cite{ref14} indicate that the
expected effect of this optical torque on an optically trapped particle
is rotation of the order of a few hertz.

To observe the mechanical effects of circularly polarized light, we
introduced a $\lambda/4$ plate into an experimental setup based on an
optical tweezers arrangement, as shown in Fig. 2. The setup differs from
the usual optical tweezers arrangement in that a laser beam is brought to
a focus below the specimen plane of a high-power objective to facilitate
trapping of absorbing particles. The absorptive material used in these
experiments is CuO powder (irregularly shaped particles approximately
1--10\,{\textmu}m in size; refractive index $n = 2.63$) dispersed in
kerosene (refractive index $n = 1.442$). The absorptivity of the CuO
particles that we used is not known; however, thin films of CuO
(50\,nm thickness) have been measured to transmit only 30\% of 1064\,nm
light~\cite{ref15}, indicating that particles of 1--10\,{\textmu}m
thickness would be highly absorbing. A drop of this mixture is placed
between a microscope slide and cover slip and then positioned in the
specimen plane of a $100\times$ high-N.A. oil-immersion objective. A
linearly polarized, Gaussian beam ($\lambda = 1064$\,nm), spatially
filtered with single-mode optical fiber, $\approx$ 20\,mW in power,
is directed into the back aperture of the objective and focused to a
diffraction-limited spot below the particle, which is then optically trapped.

\begin{figure}[htb]
\includegraphics[width=\columnwidth]{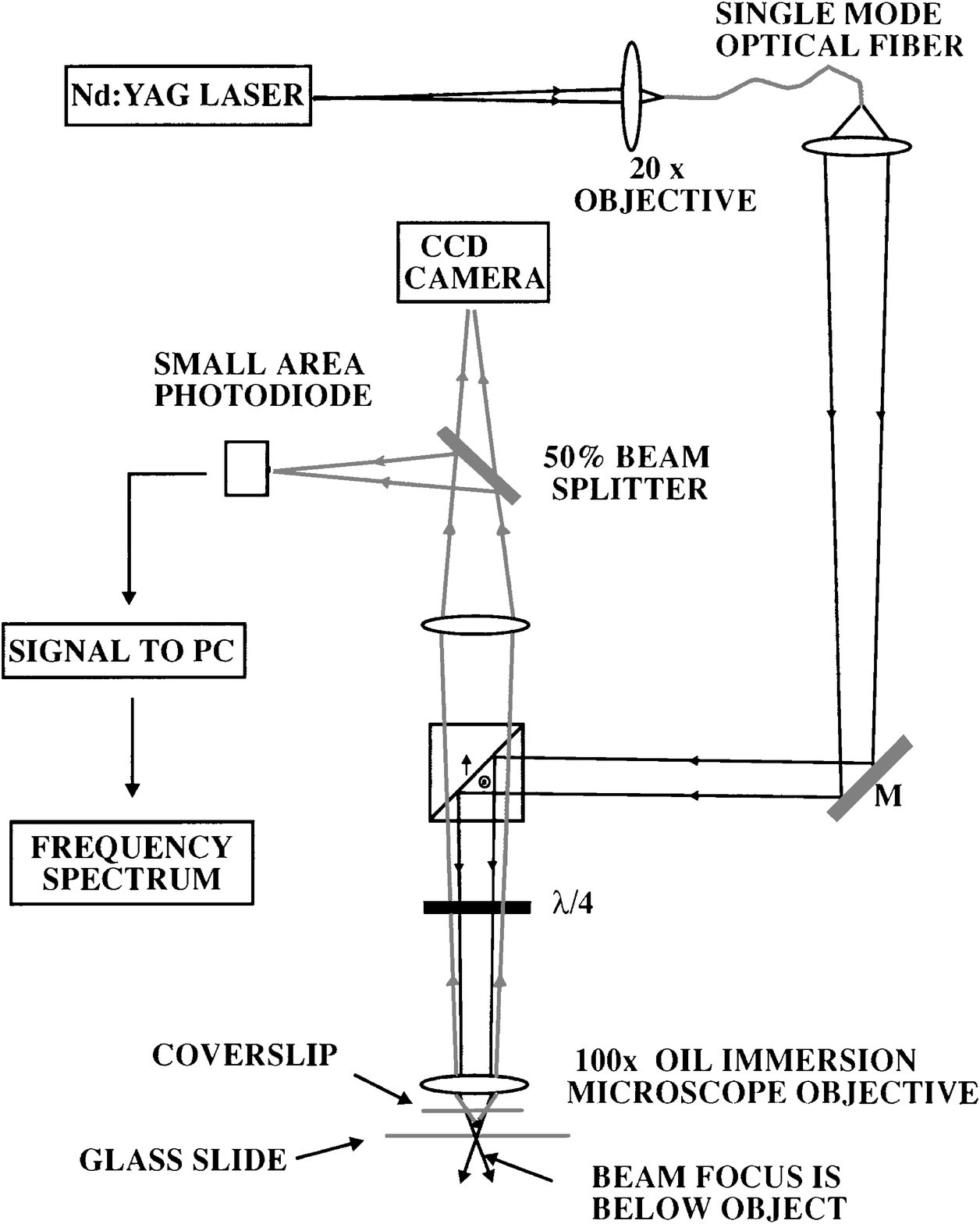}
\caption{Experimental setup for optical trapping and micromanipulation
of absorbing particles by use of a Gaussian laser beam. The particles
are trapped above the waist of the beam. The quality of the beam is
ensured to be Gaussian by optical filtering with a single-mode optical fiber.}
\end{figure}

When circularly polarized light is absorbed by a particle, by conservation
of angular momentum we expect that the particle will gain mechanical angular
momentum and thus experience a torque. The torque $\tau$ acting on a particle
of radius $r$ and absorptivity $\alpha$ trapped on the axis of a beam of
spot size $w(z)$ is given by
$\tau = (\alpha\sigma_z P/\omega) (1-\exp\{-2r^2/w^2(z)\})$, where $P$
is the beam power, $w(z)$ is the beam width, $\omega$ is the angular
frequency of the light, and $\sigma_z$ is the degree of circular polarization,
equal to $\pm 1$ for left- and right-circularly polarized light, respectively,
and 0 for plane-polarized light. For a particle rotating with angular speed
$\Omega$ in a medium of viscosity $\eta$, the drag torque is given by
$\tau_D = -8\pi\eta r^3\Omega$ and the particle rotation rate will be
constant when these torques are equal. For example, a CuO particle in
kerosene [assumed to be at 100{\textdegree}C with viscosity of
$5.5\times 10^{-4}$\,Ns/m$^2$] absorbing 10\% of a 20\,mW beam should
rotate at 10\,Hz in circularly polarized light, either clockwise or
anticlockwise about the beam axis, depending on whether the light is
left- or right-circularly polarized. This estimate is in agreement with
our observed rotation rates of 1--25\,Hz.

In our experiment we changed the beam polarization from plane to circular
through rotation of a $\lambda/4$ plate while keeping a CuO particle
trapped and observed that, on rotation of the wave plate, the particle
began to rotate at a frequency of a few hertz. All the particles that
we tested rotated in the same direction, and, on rotation of the $\lambda/4$
plate by 90\textdegree, all trapped particles
changed direction and continued to rotate. Particles did not rotate in
plane-polarized light.

The angular momentum density $\mathbf{J}$ is given by
$\mathbf{J} = (\epsilon/2\mathrm{i}\omega) \int\mathrm{d}^3r
\mathbf{E}^\star\times\mathbf{E}$, where $\mathbf{E}$ is the electric field
amplitude vector~\cite{ref16}. For elliptically polarized light produced by
passing plane-polarized light through a $\lambda/4$ plate, the electric field
vector can be written as
$\mathbf{E} = E_0 \cos\theta \hat\mathbf{j} + \mathrm{i}E_0 \sin\theta
\hat\mathbf{k}$, giving
\begin{equation}
\mathbf{J} = -\frac{\epsilon}{2\omega} E_0^2 \sin 2\theta \hat\mathbf{i},
\end{equation}
where $\theta$ is the angle between the fast axis of the $\lambda/4$ plate
and the electric field of the plane polarized incident field.

We can measure the rotation frequency of a trapped CuO particle by using
a small-area photodiode placed off center of the image of the scattered
light from the particle~\cite{ref7}, as shown in Fig. 2. To quantify the
effects of elliptically polarized light we rotated a $\lambda/4$ plate
in increments of 5\textdegree from $\theta = -45$\textdegree to
$\theta = +45$\textdegree, thus changing the polarization stepwise from
left to right circular. At each position we measured the particle's
rotation frequency, which we plot in Fig. 3 against the angle of the
$\lambda/4$ plate. Also plotted in Fig. 3 is a graph of
$f_\mathrm{max} \sin 2\theta$, which is the calculated rotation frequency
from Eq. (1), where $f_\mathrm{max}$ is the frequency of rotation in
circularly polarized light. As can be seen from the graph, the experimental
data fit the theoretical curve extremely well, confirming that the particle
rotation is in fact a result of the torque from elliptically polarized light.
Once the rotation rate for a particular particle
in circularly polarized light is known, we can control both its direction
and rate of rotation simply by rotation of a $\lambda/4$ plate.

\begin{figure}[tb]
\includegraphics[width=\columnwidth]{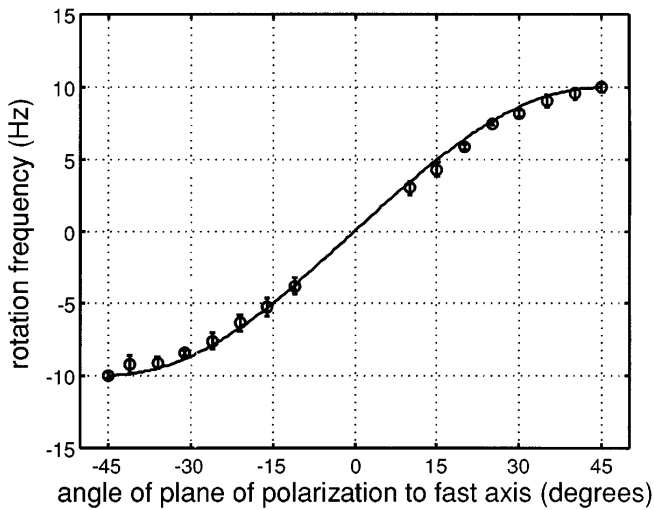}
\caption{Particle rotation as a result of the torque from elliptically
polarized light. The rotation frequency for absorbing particles trapped
in a Gaussian beam is shown as a function of $\theta$, the angle between
the fast axis of the $\lambda/4$ plate and the plane of polarization of
the incoming laser beam. The solid curve represents the expected variation
of rotation rate with $\theta$ as calculated from Eq. (1).}
\end{figure}

We have shown that it is possible to trap and manipulate strongly absorbing
microscopic particles two dimensionally without the requirement for a
doughnut beam, production of which can present substantial complications
within a conventional optical tweezers arrangement. Our investigation of
polarization effects shows that absorbing particles can be rotated without
the use of a helical doughnut beam and that, by varying
varying the ellipticity and handedness of the trapping beam polarization,
we have continuous smooth control of the rotation. This experiment provides
a direct and easily set up demonstration of the angular momentum of
elliptically polarized light.



\end{document}